\documentclass[footinbib,aps,prl,reprint,superscriptaddress,twocolumn,hidelinks]{revtex4-2}

\usepackage{graphicx}
\usepackage{rotating}
\usepackage{dcolumn}
\usepackage{amssymb}
\usepackage{amsmath}
\usepackage{bm}
\usepackage{times}
\usepackage{listings}
\usepackage{fancyhdr}
\usepackage{wrapfig}
\usepackage[caption=false]{subfig}
\usepackage{xcolor}
\usepackage{hyperref}
\hypersetup{
    colorlinks,
    linkcolor={red!50!black},
    citecolor={blue!50!black},
    urlcolor={blue!80!black}
}

\newcommand{\OO}{O}
\newcommand{\RR}{\mathcal{R}}
\newcommand{\TT}{\mathcal{T}}
\newcommand{\beq}{\begin{equation}}
\newcommand{\eeq}{\end{equation}}
\newcommand{\pa}{\partial}
\newcommand{\ed}{{\rm d}}
\newcommand{\Lie}{{\cal L}}
\newcommand{\Ord}{{\cal O}}
\newcommand{\ti}{\tilde}
\newcommand{\bra}{\langle}
\newcommand{\ket}{\rangle}
\newcommand{\ga}{\gamma}
\newcommand{\de}{\delta}
\newcommand{\De}{\Delta}
\newcommand{\La}{\Lambda}
\newcommand{\ro}{\rho}
\newcommand{\ph}{\phi}
\newcommand{\cH}{{\cal H}}

\allowdisplaybreaks

\begin{document}

\title{Infrared (in)sensitivity of relativistic effects in cosmological observable statistics}

\author{Ermis Mitsou}
\affiliation{Center for Theoretical Astrophysics and Cosmology, Institute for Computational Science, University of Zurich, CH--8057 Zurich, Switzerland}

\author{Jaiyul Yoo}
\email[]{jyoo@physik.uzh.ch}
\affiliation{Center for Theoretical Astrophysics and Cosmology, Institute for Computational Science, University of Zurich, CH--8057 Zurich, Switzerland}
\affiliation{Physics Institute, University of Zurich, Winterthurerstrasse 190, CH--8057, Zurich, Switzerland}

\author{Matteo Magi}
\affiliation{Center for Theoretical Astrophysics and Cosmology, Institute for Computational Science, University of Zurich, CH--8057 Zurich, Switzerland}

\date{\today}

\begin{abstract}
The relativistic effects in cosmological observables contain critical
information about the initial conditions and gravity on large scales.
Compared to the matter density fluctuation, some of these relativistic contributions 
scale with negative powers of comoving wave number,
implying a growing sensitivity to infrared modes.
However, this can be inconsistent with the equivalence principle 
and can also lead to infrared divergences in
the observable $N$-point statistics. 
Recent perturbative calculations
have shown that this infrared sensitivity is indeed spurious
due to subtle cancellations in the cosmological observables that have been missed in the bulk of the literature.
Here we demonstrate that the 
cosmological observable statistics are infrared-insensitive in a general and fully non-linear
way, assuming diffeomorphism invariance and adiabatic fluctuations on large scales.
\end{abstract}

\maketitle

{\it Introduction.}--- 
In cosmology, constraining model parameters with observations requires 
relating the $N$-point correlation functions of cosmological observables
to the $N$-point correlation functions of space-time fields at some 
initial hypersurface (see, e.g., \cite{PEEBL80,PEACO99,DODEL03,WEINB08B}).
Cosmological
observables~$\OO$, such as the luminosity distance, galaxy number density or cosmic microwave background
temperature, are typically a function of the observed redshift~$z$ 
of a source and its direction~$\hat{n}$ in the sky, 
together parametrizing the past light-cone. The initial condition fields, such as the comoving-gauge
curvature perturbation~$\RR$ and tensor perturbation~$\ga_{ij}$
 for minimal models, are typically considered in the 
Fourier space of the initial hypersurface, parametrized by $k^i$, 
so that statistical homogeneity and isotropy simplify their statistics. 

As an example, the two-point correlation function of the cosmological 
observables in linear perturbation theory 
for the scalar sector reads
\beq 
\label{eq:2pt}
\bra \OO(z,\hat{n}) \, \OO(z',\hat{n}') \ket = \bar{\OO}(z)\, \bar{\OO}(z')
\left[1 + \sum_{l=0}^{\infty} \Lie_l(\mu) \, C_l(z,z')\right]
 \, ,  
\eeq
where $\Lie_l$ are the Legendre polynomials with cosine angle 
$\mu:=\hat{n}\cdot\hat{n}'$,
\beq \label{eq:CofP}
C_l(z,z') = \int d\ln k ~\TT_l(z; k) \, \TT_l(z';k) \,\Delta^2_{\RR}(k) ~,
\eeq
is the angular power spectrum of the cosmological observables,
and $\Delta^2_{\RR}:=k^3P_\RR(k)/2\pi^2\propto k^{n_s-1}$ 
is the dimensionless power spectrum of the primordial fluctuations~$\RR$ with
spectral index~$n_s$.
All the information about the evolution of the universe and the light 
propagation to the observer is then contained in the kernel or the transfer 
function~$\TT_l(z;k)$. 
The generalization of this relation in Eq.~\eqref{eq:CofP}
to higher-order perturbation theory and other $N$-point
statistics retains the same 
qualitative form of a convolution with kernels that depend on multipole 
indices, redshifts and initial-hypersurface Fourier wave-vectors \footnote{The Fourier coordinates $k^i$ considered here should not be confused with the ones observers use to map the sky, as the two types are equivalent only in the Newtonian limit. Indeed, the ``observer's Fourier space'' is the dual of Cartesian coordinates inferred from the data $(z,\hat{n})$ through some fiducial cosmology and thus parametrizing the past light-cone, {\it not} a constant-time hypersurface. Consequently, in this observational Fourier space large scales correspond to early times, meaning that there is either no well-defined infrared limit (finite particle horizon), or that limit is trivial since structure is absent in the distant past.}. 

The Fourier dependence of these kernels can then be organized in decreasing orders of spatial derivatives, i.e. the derivatives that act on the 
primordial fields~$\RR$ 
before taking the ensemble average. This ordering makes particular sense,
since large-scale surveys have historically progressed by probing from
 small to larger scales (see, e.g., \cite{PEBAET01,EIZEET05}),
so higher-order derivative terms 
have been most relevant in practical situations. 
At the level of the kernels, such as $\TT_l$ in Eq.~\eqref{eq:CofP}, this corresponds to ordering the terms by $(\cH/k)^{n\geq 0}$ factors with respect to the dominating contributions on small scales, or the matter density 
fluctuation~$\de_m\sim k^2\RR$, where~$\cH$ is the conformal Hubble parameter.
While the lowest-order approximation $\cH/k \to 0$ has been accurate so far, 
the upcoming surveys 
\cite{LSST04,SKA09,EUCLID11,WFIRST12,DESI13,SPHER14} 
will be sensitive to the contributions $\sim (\cH/k)^{n>0}$.
These terms contain information about the initial conditions in the early
Universe and relativistic effects, both in field equations and the light propagation between source and observer. 

If the integrand leading to the observable $N$-point 
statistics contains low enough powers of~$k$ (or $n>0$),
then the result can be {\it dominated} by infrared mode contributions, 
i.e. with their influence growing indefinitely as $k \to 0$. 
In general the observer-source system is in free-fall and of finite extent, so it should gradually {\it lose} sensitivity to fluctuations of wavelength
longer than the separation between the source and the observer,
because the infrared limit that is a uniform gravitational potential and force is equivalent to an accelerated coordinate system (equivalence principle).
In fact, in the most extreme case the Fourier integral in Eq.~\eqref{eq:CofP}
can even be 
{\it infrared-divergent}, 
which is even more unphysical for observations involving a finite space-time patch. As it turns out \cite{SASAK87,FUSA89,BODUGA06,YOFIZA09,YOO10,CHLE11,BODU11,JESCHI12,BEDUET14,UMCLMA14a,YOZA14,BEMACL14,DIDUET14,BOCLET15a,KOUMET18,MAYO22},
the cosmological observables of most interest do contain relativistic effects of low enough $k$-powers ($n>0$) for this to happen. 
Importantly, the current standard approach to modeling the cosmological
observables in the literature leads to infrared-divergences.
As a result, the bulk of the works in the literature offers an {\it inaccurate} 
modeling of the infrared contributions to the observable $N$-point
statistics on large scales,
precisely the region of interest for upcoming observations. 

In this Letter we provide a general and fully non-linear proof that the statistics of cosmological observables remain insensitive to infrared modes when relativistic effects are taken into account, assuming adiabatic fluctuations on large enough scales.
If one expands the full $k$-dependence of the kernels, such as 
$\TT_l$ in Eq.~\eqref{eq:CofP}, then the lowest powers of $k$ from 
individual relativistic effects {\it cancel each other out exactly}. 
The precise {\it total} impact of the relativistic effects in the cosmological observables
is therefore more subtle than some power 
increase $\sim (\cH/k)^{n>0}$ at low~$k$ (see \cite{GRSCET20}
for the infrared behaviors in Fourier space, which are still non-trivial).
A sufficient set of conditions for this low $k$-power cancellation are diffeomorphism invariance and adiabatic fluctuations at large enough scales, both of which hold in the standard cosmology, or the $\La$CDM model.
Our conclusions are supported by several explicit perturbative computations 
at linear order 
\cite{JESCHI12,BIYO16,BIYO17,SCYOBI18,GRSCET20,BAYO21,CADI22},
to which we recently added the first detailed non-linear example 
\cite{YOGRMI22}. Our present demonstration, however, does not restrict to a particular observable and applies to {\it all orders in perturbation theory}.

{\it Infrared (in)sensitivity of the cosmological observables in linear
  theory.}--- 
In the Newtonian theory the cosmological observables only depend on the 
gravitational potential~$\ph$ through its second spatial derivatives 
$\pa_i \pa_j \ph \propto \pa_i \pa_j \RR$~, 
i.e. the matter density contrast~$\de_m$ 
(trace part) and the tidal tensor (traceless part). 
In the relativistic theory, however, some contributions to the cosmological observables 
are directly proportional to the metric fluctuations
$h_{\mu\nu} \propto \RR$ 
(e.g., the Sachs-Wolfe effect \cite{SAWO67}), 
including gravitational waves~$\ga_{ij}$,
so they contribute with lower powers of $k$ in the kernels of the observable 
$N$-point statistics \cite{SASAK87,FUSA89,BODUGA06,YOFIZA09,YOO10,CHLE11,BODU11,JESCHI12,BEDUET14,UMCLMA14a,YOZA14,BEMACL14,DIDUET14,BOCLET15a,KOUMET18,MIYOET20,MAYO22}.

At the linear order in perturbations, the only non-trivial relation 
for the observable $N$-point statistics is Eq.~\eqref{eq:CofP}.
In the Newtonian theory, $\bra \pa^2 \ph \, \pa^2 \ph \ket \sim k^4 P_{\RR}(k)$, 
so $\TT_l \sim\Ord(k^2)$, 
while the relativistic effects contribute at $\TT_l \sim \Ord(k^{0\leq n<2})$, and 
$\bra \ph \ph \ket \sim P_{\RR}(k)$ at worst. 
Incidentally, the local primordial non-Gaussianity in galaxy clustering 
also induces such low-derivative terms through the bias model 
\cite{DADOET08,MAVE08,SLHIET08}, i.e., a term 
$\bra \de_m \ph\ket \sim k^2 P_{\RR}(k)$. Now since $P_{\RR}(k) \sim k^{n_s-4}$ at low $k$, the most 
infrared-sensitive $\Ord(k^0)$ contributions in $\TT_l$ lead to 
an infrared-divergent integral in Eq.~\eqref{eq:CofP} 
for non-blue spectra $n_s \leq 1$, as observations currently favor 
(see, e.g., \cite{PLANCKcos15,PLANCKcos18}).
At the level of the angular power spectrum $C_l(z, z')$, 
this divergence is relatively harmless, because it arises only 
in the monopole~$C_0$ \cite{YOFIZA09,YOO10,CHLE11,BODU11,JESCHI12,MIYOET20}
(and also in the quadrupole~$C_2$ 
if one includes tensor modes~$\ga_{ij}$ \cite{ADDUTA16}), 
which is usually ignored in literature.
However, since all multipoles contribute to the correlation function in 
Eq.~\eqref{eq:2pt}, the presence of infrared divergences in {\it any} multipoles
leads to inconsistencies in the observable two-point correlation function.
Moreover, it was shown \cite{YOMIET19,YOMIET19T,YOO20}
that ignoring the monopole fluctuation~$C_0$ in the CMB or supernova
observations because of such divergences {\it biases} 
the cosmological parameter estimation and underestimates their uncertainties.

This issue was resolved in a series of recent works for the luminosity distance \cite{BIYO16,BIYO17}, galaxy clustering \cite{JESCHI12,SCYOBI18,GRSCET20,CADI22} and cosmic microwave background \cite{BAYO21}, 
which showed that with a fully gauge-invariant expression for the cosmological
observables
the infrared-sensitive contributions {\it actually cancel each other out}.
 More precisely, by assuming linear Gaussian $\La$CDM and comoving observer/sources with the free-falling matter fluid, and by expanding the kernel~$\TT_l(k)$ 
in~$k$ one finds that all  terms $\propto k^{0\leq n<2}$ 
sum up to zero. 
Thus, the cosmological observable $\OO(z,\hat{n})$ in the relativistic
theory is {\it just as infrared-insensitive as its Newtonian counterpart} 
$\TT_l\sim \Ord(k^2)$.
Importantly, for this cancellation to occur 
{\it the fluctuations must be adiabatic in the infrared limit} 
\cite{BAYO21,CADI22}, which is the case in $\La$CDM. Indeed, setting $\lim_{k\to 0} k^n \TT^2_l(k) = 0$ for $n \in \{ 0,1,2,3 \}$ one finds the equations satisfied by Weinberg's adiabatic mode \cite{WEINB03}, i.e., the $k \to 0$ limit of adiabatic solutions.  

Another important condition for this cancellation to take place is that {\it one must consistently include all the relativistic
contributions to the observable} $\OO(z,\hat{n})$, so that the coordinate-independence of the theoretical expression is satisfied
\cite{JESCHI12,BIYO16,BIYO17,YODU17,SCYOBI18,GRSCET20,BAYO21,CADI22}.
The general form of a cosmological observable at the linear order 
in perturbations is
\beq \label{eq:Dols}
\OO(z,\hat{n}) = \bar{\OO}(z)\left[1 + X_{\rm o}(\hat n) 
+ X_{\rm los}(z,\hat{n}) + X_{\rm s}(z,\hat{n})\right] \, ,
\eeq
where $\bar{\OO}$ is the background contribution, $X_{\rm o}$ represents quantities at the observer point, $X_{\rm los}$ field fluctuations integrated over the background line-of-sight path and $X_{\rm s}$ fields evaluated at the source 
point \footnote{Note that this separation is somewhat arbitrary,
  as one can move some terms around by integrating by parts the line-of-sight
  terms. Moreover, although each individual contribution can be
  gauge-dependent, the sum $\OO(z,\hat{n})$ is invariant under {\protect
  \it any} coordinate transformation at each order, i.e., both the usual gauge
  transformations of cosmological perturbation theory and more general elements
  of the diffeomorphism group \cite{MIYO20}.}.
 In the standard practice the ``observer terms'' $X_{\rm o}$ are simply discarded (see below), thus yielding infrared-sensitive results. However, since $X_{\rm o}$ has 
{\it no} position 
dependence~$(z,\hat n)$, it can only contribute up to the first few
multipoles~$C_s$, where $s=0,1,2$~is the highest spin of the fields/quantities 
appearing in $X_{\rm o}$, 
 which is why the spurious infrared divergences only occur for the first few multipoles 

{\it Infrared (in)sensitivity in non-linear theory.}--- 
At the non-linear level the field combinations appearing in the integrals of $\OO(z,\hat{n})$ generalize as follows: the dominant ones at small scales are of the form $\sim h^n \pa^2 h$, while the more infrared-sensitive (lower-derivative) ones are $\sim h^n$, $h^n \pa h$ and $h^n \pa h \pa h$, i.e. all possible combinations with up to two spatial derivatives (from light propagation and field equations). 
The first example of infrared insensitivity at non-linear order was shown for the observed galaxy number density bispectrum [36]: in the squeezed equal-redshift configuration 
$\bra \OO(z,\hat{n})\, \OO(z,\hat{n})\, \OO(z,-\hat{n}) \ket$ 
the leading-order relativistic effect contributions in the small momentum variable $k_l$ cancel out. The computation is long and non-trivial, but the busy reader can still have a look at Eq.~(4.32) of \cite{YOGRMI22}, where each contribution to the signal is labelled by its physical interpretation and discussed. One can then see explicitly the cancellation between light-propagation effects and in particular the necessity of including the observer terms for this to happen.

An important aspect of non-linear perturbation theory is that, unlike the linear case, the relativistic effects at the observer position
affect {\it all} multipoles of {\it all} observable multi-spectra. Indeed, at non-linear order there are also cross-type terms in $\OO(z,\hat{n})$, e.g., of the form $X_{\rm o} \times X_{\rm s}(z,\hat{n}) \subset \OO^{(2)}(z,\hat{n})$, 
thus contributing terms like $\bra [X_{\rm o} X_{\rm s}(z,\hat{n})] \, [ X_{\rm o} X_{\rm s}(z',\hat{n}') ] \ket$ in the 2-point function and $\bra [X_{\rm o} X_{\rm s}(z,\hat{n})] \, X_{\rm s}(z',\hat{n}') \, X_{\rm s}(z'',\hat{n}'') \ket$ in the 3-point function. {\it There is a priori no reason why the corresponding contributions would be smaller than the ones without any observer dependence, especially at large scales}. The observer term issue is therefore {\it  not at all trivial} non-linearly, yet ignored in most
 existing computations of the bispectrum.
Although the non-linear realm currently lacks a more general rigorous exploration, i.e. including all terms at a given order, one can still expect cancellations of the lowest powers of the wave-number dependencies that are expected from individual relativistic effects. Indeed, as we will show below, the underlying reason is the diffeomorphism symmetry, which is independent of the perturbative order.

{\it General proof of infrared insensitivity.}--- 
Let us now demonstrate the infrared insensitivity of relativistic effects in the 
cosmological observables. 
We start by noting that cosmological observations
involve a {\it finite} patch of space-time -- the one containing the 
observer-source system -- whose typical length-scale we denote by $L_z$. Let us then choose a reference scale $L \gg L_z$ and split the metric perturbation $h_{\mu\nu}$ into long and short wavelength contributions $h_{\mu\nu} \equiv h_{\mu\nu}^{\rm L} + h_{\mu\nu}^{\rm S}$, in some given coordinates. 
One possibility for this split is the Gaussian coarse-graining
\beq \label{eq:hLreal}
h_{\mu\nu}^{\rm L}(t,x) := \frac{1}{(2\pi)^{3/2} L^3} \int \ed^3 y
 \exp \left[ - \frac{| x-y |^2}{2 L^2} \right] h_{\mu\nu}(t,y)  \, ,
\eeq
or in Fourier space,
\beq \label{eq:hLFourier}
h_{\mu\nu}^{\rm L}(t,k) \equiv \exp \left[ - \frac{1}{2} \, L^2 k^2 \right] 
h_{\mu\nu}
(t,k)  \, .
\eeq
As a result, the short-mode metric fluctuation is
\beq \label{eq:hSFourier}
h_{\mu\nu}^{\rm S}(t,k) \equiv h_{\mu\nu}(t,k)-h_{\mu\nu}^{\rm L}(t,k)
 \sim \Ord(k^2 h_{\mu\nu}) \, , 
\eeq
so the infrared-sensitive $\Ord(k^{0\leq n < 2})$ contributions to the 
cosmological observable $\OO(z,\hat{n})$ would disappear if $h_{\mu\nu}^{\rm L} = 0$.  

By construction $h^{\rm L}_{\mu\nu}$ varies very little within a distance $L_z$ from the observer world-line $x^i_{\rm o}(t)$,
since we chose $L \gg L_z$. We can therefore approximate the long-mode
metric fluctuation by its first-order expansion around the world-line
\beq \label{eq:hLexpand}
h^{\rm L}_{\mu\nu}(t,\bm{x}) \approx f_{\mu\nu}(t) + \left[ x^i - x^i_{\rm o}(t) \right] f_{i,\mu\nu}(t) \, . 
\eeq 
Note that the corrections $\Ord (\De x^2/L^2)$ can be made arbitrarily small by increasing $L$ and thus negligible for the purpose of computing 
$\OO(z,\hat{n})$, as it only involves fields within a distance $L_z$ from the world-line. Put differently, using a large, yet finite $L \gg L_z$, the approximated long-mode metric in Eq.~\eqref{eq:hLexpand} yields an observable that is indistinguishable from the exact one.
Incidentally, note that the second-order derivative terms 
$\sim h^n \pa^2 h$ are precisely those that vanish in Eq.~\eqref{eq:hLexpand}, 
while the infrared-sensitive ones ($h^n$, $h^n \pa h$ and $h^n \pa h \pa h$) do not. 

The central observation now is that, under some conditions discussed below, the precise form in Eq.~\eqref{eq:hLexpand} can be removed by a {\it residual} large coordinate transformation \cite{WEINB03,CRNOSI12,HIHUKH12,CRNOET13},
i.e., one that preserves the chosen gauge and therefore {\it the expression of the observable in terms of the perturbation fields}. Since cosmological observable relations are {\it fully} coordinate-independent \cite{MIYO20,MIYO22b}, they are invariant under this manipulation of a coordinate transformation, so the result of their computation must be the same when $h_{\mu\nu}^{\rm L} = 0$. 
Given that this is the part of $h_{\mu\nu}$ that carries the infrared-sensitive
terms $\Ord(k^{0\leq n < 2})$ 
in the integrals of the observable $\OO(z,\hat{n})$, we conclude that the corresponding infrared-sensitive contributions for the observable $N$-point 
statistic must be absent \footnote{Note that one can always eliminate $h^{\rm L}_{\mu\nu}$ in Eq.~(7), i.e. for all values of $f_{\mu\nu}$ and $f_{i,\mu\nu}$, by a coordinate transformation such as the one to conformal Fermi coordinates \cite{DAPASC15a}, but the resulting expression of the observable would be different in terms of the perturbations in the new coordinates, hence the need for a {\it residual} gauge transformation in our argument.}.

In the typical gauges chosen in the literature one has 
$h_{\mu\nu}^{\rm L} \neq 0$, so the absence of low $k$-powers can only occur 
through {\it cancellations among infrared-sensitive terms}. Since the 
argument above relies on the diffeomorphism symmetry, one must 
include {\it all the relativistic contributions} in the cosmological observables
at a given order to maintain its coordinate-independence, and this is not
possible if some contributions -- for example at the observer position -- are
ignored.
Intuitively, including those terms at the observer position in $\OO(z,\hat{n})$ can then be understood as consistently implementing the {\it finite} observer-source separation $L_z$ that sets a reference scale with respect to which infrared modes are filtered. 

{\it Required conditions.}--- The long-mode part given by Eq.~\eqref{eq:hLexpand} can be eliminated through a residual coordinate transformation 
\beq \label{eq:htoth}
h_{\mu\nu} \to \ti{h}_{\mu\nu} \, , \hspace{1cm} \ti{h}_{\mu\nu}^{\rm L} = 0 \, ,
\eeq
only for specific profiles $h_{\mu\nu}^{\rm L}$, i.e. specific functions $f_{\mu\nu}(t)$ and $f_{i,\mu\nu}(t)$. Solutions that become ``pure-(residual)-gauge'' in the infrared limit are known as ``adiabatic modes'' \cite{WEINB03} (see \cite{PAJA18} for an up-to-date systematic exploration). Equivalently, adiabatic modes can be generated by acting on purely short-mode solutions with specific diffeomorphisms that preserve the gauge. At the linear level, where all modes decouple, this means that they can be generated by acting with some coordinate transformation on the background solution and in the matter sector this leads to the relations 
\beq \label{eq:adiab}
\frac{\de \ro^a}{\dot{\bar{\ro}}^a} = \frac{\de p^a}{\dot{\bar{p}}^a} = \frac{\de \ro^{\rm tot}}{\dot{\bar{\ro}}^{\rm tot}} = \frac{\de p^{\rm tot}}{\dot{\bar{p}}^{\rm tot}} \, ,
\eeq
which is thus consistent with the usual definition of ``adiabatic modes'' in linear cosmological perturbation theory. Given Eq. \eqref{eq:hSFourier}, an adiabatic $h_{\mu\nu}^{\rm L}$ means that any non-adiabatic contribution in $h_{\mu\nu}$ must decay at least as fast as $\sim k^2$ in the infrared. 

An important subtlety is that acting with a large diffeomorphism can also generate unphysical solutions. The typical example in linearized general relativity
(GR)
\cite{WEINB03} is the constant dilatation $x^i \to (1+\lambda) x^i$ in the Newtonian gauge which transforms the potentials as $\ph \to \ph + \lambda$ and $\psi \to \psi$.  The resulting solution no longer satisfies $\ph = \psi$ at $k = 0$ (assuming no anisotropic stress), which is allowed since that equation is actually $\pa^2 \ph = \pa^2 \psi$. Adiabatic modes therefore include the constraint of being smooth $k \to 0$ limits of finite-$k$ solutions and, as a result, their profile depends on the equations of motion of the theory. 
For instance,
if we consider  a modified gravity theory, then there would still exist
adiabatic modes, but the corresponding functions~$f_{\mu\nu}(t)$ and
$f_{i,\mu\nu}(t)$ of Eq.~\eqref{eq:hLexpand}  could be different from those
in GR, since the
equations of motion would be different. However, most modified gravity
theories contain GR as a subset of solutions (e.g., Horndeski theory
solutions with constant scalar), so in that case one would at least have the
adiabatic modes of GR.

Let us now mention the most relevant adiabatic modes for GR in the context of cosmology. There is the already mentioned scalar Weinberg adiabatic mode \cite{WEINB03,CRNOET13}, which is constant in space, but also its tensor and conformal group generalization that includes a linear $\sim x^i$ term \cite{CRNOSI12,HIHUKH12,CRNOET13}. This is exactly the spatial dependence we need to reproduce 
the metric in Eq.~\eqref{eq:hLexpand} for cosmological applications. More precisely, it has been shown that one can generate the Newtonian-gauge scalar components \cite{WEINB03,CRNOET13} and both scalar and tensor components in the comoving-gauge \cite{CRNOSI12,HIHUKH12}. There are two subtleties in the presence of tensor modes: the linear $\sim x^i$ scalar profile can be generated only if there are no short-mode tensor modes $\ti{\ga}_{ij} = 0$ \cite{CRNOSI12,HIHUKH12} and one can only generate $\ga_{ij}^{\rm L}$ to linear order in that field \cite{HIHUKH12}.

The condition that only the adiabatic scalar degree of freedom is active above some scale is satisfied in $\La$CDM, since this property is satisfied in the
initial conditions and preserved through evolution non-linearly at scales above the largest sound horizon among the involved propagating fields. Thus, choosing~$L$ above the largest sound horizon and $L_z$, we conclude that {\it in $\La$CDM all cosmological observables are infrared-insensitive to all orders in perturbation theory.}

The condition of adiabaticity at the linear level was already pointed out in \cite{BAYO21} through its standard form in Eq.~\eqref{eq:adiab} and in \cite{CADI22} through the appearance of the equations describing Weinberg's adiabatic mode. Intuitively, deviations from adiabaticity correspond to more than one physical reference frame, such as the case of two distinct bulk flows influencing the observer-source system. Even if one eliminates their average effect (adiabatic mode) through the equivalence principle, their coordinate-independent difference (entropy mode) will remain (see, e.g., \cite{MAVABR13,BRGRET18}).
The complications that arise in the presence of tensor modes, as described above, could also be understood as a deviation from adiabaticity, i.e. an extra physical reference that obstructs or remains after the elimination of a constant force.    

Finally, so far we have implicitly assumed that all relativistic effects are expressible in terms of $h_{\mu\nu}$, which is the case only if one assumes free-falling observer and sources. In the explicit computations 
\cite{JESCHI12,BIYO16,BIYO17,SCYOBI18,GRSCET20,BAYO21,CADI22} both were taken to follow the matter fluid, itself being in free-fall, which is accurate at the large scales of interest. However, this free-fall assumption for the observer and sources is not necessary, thanks to the requirement of adiabaticity at large enough scales. Independently of whether these actors are subject to non-gravitational forces, or follow different non-free-falling fluids, the important point is that their dynamics at cosmological scales are ultimately determined by some of the available fields in the theory of interest. These fields will then enter in the expression for the observable through the observer/source perturbations. However, since only the single (adiabatic) degree of freedom survives at the largest scales, the infrared part of the extra fields will be expressible in terms of $h_{\mu\nu}$ as well.

{\it The conceptual issue with observer terms}--- We argued that consistent statistics require the inclusion of observer terms in $\OO(z,\hat{n})$. However, for their contribution to take the required form, i.e. proportional to primordial spectra \eqref{eq:CofP}, observer terms must be expressed as stochastic fields evaluated at the observer position. This is achieved by relating the observer quantities to the underlying fields that govern the dynamics (e.g. $\bm{v}_{\rm o} = \bm{v}_m(t_{\rm o}, \bm{x}_{\rm o})$ for a comoving observer). Thus, the observer adapts to each universe realization in the ensemble average, as the sources do.

That approach, however, raised objections in private communications and workshops, which revealed more concretely why observer terms were ignored for so long in the literature. Critics would point out that the observer terms $X_{\rm o}$ should not be considered as stochastic, as some of them can in principle be measured with local experiments, in contrast to the rest of the light-cone information, which is available only in a model-dependent way. 

Importantly, there can be no ``right'' and ``wrong'' approach, since these are just two different prescriptions for doing statistics with our single universe. Which part of the information we model as deterministic or stochastic is a matter of choice, similarly to the question of prior knowledge in Bayesian statistics. Thus, the fact that some terms in $X_{\rm o}$ {\it could} be measured directly in principle does not force one to fix that value. The question therefore is: which prescription is the most convenient, most effective, or even simply computable? 

The local measurements required in the fixed-observer prescription are dominated by the highly non-linear local dynamics, which one does not resolve within perturbation theory, so their inclusion in this context would be ambiguous. Moreover, although some quantities at the observer are measurable (e.g. our relative velocity to the CMB), most of them are not (e.g. the scalar potentials), as they cannot be expressed in terms coordinate/frame-independent quantities (e.g. local curvature invariants). In contrast, the ``statistical-observer'' prescription advocated here requires no extra input and is straightforwardly computable. The condition of a comoving observer (and sources) with matter is accurate at the scales one resolves within perturbation theory and it is also an unambiguous, physical condition. Moreover, the consistency of this approach was demonstrated in \cite{MIYOET20}, where we showed that, if we proceed without privileging the observer position at any step, then the variance of unbiased estimators is {\it exactly} cosmic variance, i.e. the usual information bound in cosmology.

{\it Discussion.}--- We saw that infrared insensitivity is guaranteed {\it if} non-adiabatic fluctuations decay at least as fast as $\sim k^2$ at low $k$. This condition is intriguing, because one could a priori think of instances where it does not hold, yet breaking it too much would lead to infrared-divergent observables -- clearly an unphysical result for a computation involving a finite space-time patch. For instance, this happens at the linear level if non-adiabatic contributions do not decay ($\TT_l \sim k^0$) and $n_s \leq 1$. One is therefore led to wonder whether there exists any mechanism that can produce non-adiabatic
fluctuations that survive at $k = 0$.
Lacking any definite answer on this matter, we can at least say that this does seem at odds with the underlying framework of cosmological perturbation theory, as one would expect homogeneity and isotropy at $k \to 0$, thus leaving only room for solutions that are coordinate artefacts in that limit, i.e. adiabatic solutions.

\vfill

\begin{acknowledgments}
We are grateful to the participants of the workshop ``General Relativistic effects in observing the Large Scale Structure of the Universe'' (June 2022, Porto) for useful discussions. JY acknowledges useful discussions with Robert Brandenberger.
EM is supported by a Forschungskredit Grant of the University of Zurich (grant FK-21-130) and JY is supported by the Swiss National Science Foundation and a Consolidator Grant of the European Research Council.
\end{acknowledgments}

\bibliography{ms.bbl}

\end{document}